\documentstyle[12pt]{article}
\begin{document}
\title{CONSEQUENCES OF A QUANTIZED SPACETIME MODEL$*$}
\author{B.G. Sidharth\\ Centre for Applicable Mathematics \& Computer Sciences\\
B.M. Birla Science Centre, Hyderabad 500 063 (India)}
\date{}
\maketitle
\footnotetext{$^*$ Based on an invited talk at the Department of Atomic Energy,
Government of India, Symposium on High Energy Physics, 2000.\\
E-mail:birlasc@hd1.vsnl.net.in}
\begin{abstract}
A recent extended particle model is discussed, which lead to some interesting
consequences in cosmology, neutrino astrophysics and low dimensional and low
temperature statistics, some of which have since been verified.
\end{abstract}
\section{The Model}
Recent studies go beyond the spacetime points and point particle description
of usual Quantum Mechanics and Quantum Field Theory. These include String
Theory\cite{r1,r2,r3}, El Naschie's theory of Cantorian
spacetime and transfinite heterotic String Theory \cite{r4,r5} and
the author's theory of stochastic quantized spacetime \cite{r6,r7}.\\
The starting point of this model was the fact that the purely classical Kerr-
Newman Black Hole describes the electron's field including the purely Quantum
Mechanical $g=2$ factor. However there are two inexplicable features. The first
is the naked singularity of the Kerr-Newman Black Hole, if it is to represent
an electron. The second is the complex coordinate shift used by Newman in
deducing the Kerr-Newman metric, which Newman himself could not explain. It was
pointed out\cite{r8,r9,r10} that there is a spacetime cut off, at the Compton
scale $(l, \tau)$ exactly as in the case of Dirac's non Hermitian position
operator for the electron. From the Uncertainity Principle itself it follows
that spacetime points are meaningless as they imply infinite momenta
and energy. Rather spacetime intervals are meaningful.  Infact the zitterbewegung
is symptomatic of the breakdown of physics within this scale. It is this
spacetime cut off that fudges the naked singularity and infact averts what
Wheeler has described as the greatest crisis of physics.\\
Indeed Newman's complex coordinate shift can be seen to represent the spacetime
cut off: For if
$$x \to x + \imath l$$
then the plane wave goes over to
\begin{equation}
e^{\frac{\imath p_x}{\hbar}} \to e^{\imath / \hbar p_x}\cdot e^{-\frac{pl}{\hbar}}\label{e1}
\end{equation}
We can see from (\ref{e1}) that as a result of the complex shift the plane
wave is truncated, by the constraint
\begin{equation}
pl \sim \hbar\label{e2}
\end{equation}
the relation (\ref{e2}) can be seen to be the momentum distance Uncertainity relation, as noted
above.\\
We will now review some of the consequences of this formulation, and point out
tht many of these have since been experimentally or observationally confirmed.
\section{Consequences}
{\large \bf 1. Cosmological Considerations}\\
If we use the fact that there would be a fluctuational creation of $\sqrt{N}$
particles within the minimum Compton cut off time $\tau$, where $N$ is the
total number of particles, typically pions, in the universe at a given epoch,
we deduce the following relations \cite{r11,r12,r13}
\begin{equation}
m = \left(\frac{\hbar^2 H}{Gc}\right)^{\frac{1}{3}}\label{e3}
\end{equation}
\begin{equation}
\frac{dR}{dt} = HR\label{e4}
\end{equation}
where $H$ is the Hubble constant,
\begin{equation}
\Lambda \le 0 (H^2)\label{e5}
\end{equation}
$\Lambda$ being the cosmological constant,
\begin{equation}
G \propto \frac{1}{T}\label{e6}
\end{equation}
where $T$ is the age of the universe.
Equation (\ref{e3}) is known empirically, and has been described by Weinberg
to be mysterious. Here it follows as a consequence. Equations (\ref{e4}) and
(\ref{e5}) show that the universe would continue to expand for ever (infact in this case,
with decreasing density) and possibly also accelerate. This was subsequently
confirmed by the observations of distant supernova\cite{r14}.\\
It must be mentioned that till these recent observations were made, it was believed that
the expansion of the universe would be subsequently reversed. It may also be
mentioned that the above model does not need to invoke dark matter which in
any case has not been detected. The relation (\ref{e6}) shows that the universal
constant of gravitation varies with time, as in a few other cosmological
models. This is as yet an undecided matter.\\
{\large \bf 2. Anomalous Statistics}\\
Following these arguments it was shown that the divide between Fermi-Dirac
statistics and Bose-Einstein statistics is not so water tight\cite{r15,r16,r12,
r17,r18,r19,r20,r21}.\\
This fact has lead to some interesting results. One was that the Neutrino would
have a mass given by \cite{r22,r23}.
\begin{equation}
m_\nu \sim 10^{-8} m_e\label{e7}
\end{equation}
and infact there would be a "weak" electric charge given by
\begin{equation}
g^2_\nu /e^2 \sim 10^{-13}\label{e8}
\end{equation}
Subsequently the superkamiokande experiments showed exactly such a neutrino
mass as given in (\ref{e7})\cite{r24}.\\
Yet another interesting result was the super conductivity type of behaviour of
electrons in low dimensions, in particular one and two dimensions. These would be
idealizations of thin wires and thin films. It was shown\cite{r15}
that in a thin wire which can be approximated as a one dimensional object, the
electrons would behave as if they were below the Fermi temperature, whatever
be the temperature. It is interesting that recent observations with nano tubes
do indeed confirm such features\cite{r25,r26,r27}.\\
Further it was shown that\cite{r17,r20,r21} an approximately
mono energetic collection of Fermions would show bosonization effects and
vice versa, including a Bose-Einstein type of condensation a little above
absolute zero. These need to be examined experimentally.\\
{\large \bf 3. Quarks and Monopoles}\\
It was shown that electromagnetism was the result of the double connectivity
of the spin half electron, brought out by the fact that well outside the
Compton wavelength it is the positive energy solutions which are invariant
under reflection that predominate \cite{r8,r9}. From these considerations it was possible
to deduce the well known gravitational force - electromagnetism ratio
\begin{equation}
Gm^2 / e^2 \sim 10^{-40}\label{e9}
\end{equation}
However, it was argued that as we approach the Compton wavelength the double connectivity
or three dimensionaity of space breaks down as we begin to encounter predominantly
the negative energy components of the Dirac bi spinor
(with opposite parity), and this was shown to
explain the fractional charges of the quarks, and also provide an order of
magnitude estimate for their masses as also their handedness\cite{r28,r29}. This
would also explain why free quarks are never seen in nature. A similar
explanation holds for monopoles \cite{r30,r31} (Cf. Appendix).\\
It is interesting that from the above considerations, using relations like
(\ref{e8}) and (\ref{e9}), we get the well known ratio of all
interactions\cite{r32}
$$g^2_{strong} : g^2_{em} : g^2_{wk} : g^2_{grav} \sim 1:10^{-3} : 10^{-15} : 10^{-40}$$
{\large \bf APPENDIX}\\ \\
In \cite{r8,r9} it was argued that one could get a reconciliation
between Quantum Mechanics, Electromagnetism and Gravitation, from the following
consideration:\\
We use the well known fact that the Dirac four spinor which describes
the electron has the negative energy spinors $\chi$ and the positive energy
spinors $\theta$ and that as we approach the Compton wavelength, it is the
negative energy spinors which dominate, and further under reflections, $\chi$
behaves like the pseudo spinor,
\begin{equation}
\chi \to -\chi\label{e10}
\end{equation}
It was pointed out that this leads to a coviariant derivative,
\begin{equation}
\frac{\partial \chi}{\partial x^\mu} \to \frac{\imath}{\hbar}\left[\frac{\hbar}
{\imath} \frac{\partial}{\partial x^\mu} + \imath N A^\mu \right] \chi\label{e11}
\end{equation}
where
$$A^\mu = \hbar \Gamma^{\mu \sigma}_{\sigma} = \hbar \frac{\partial}{\partial x^\mu}
log (\sqrt{|g|})$$
(and $N = 1$ is the weight of $\chi$ which shows up as a densiity).\\
We would like to point out that this is exactly the circumstance for the Dirac
monopole.\\
What this means is that it is the region at or within the Compton wavelength where
the negative energy spinors predominate that shows up as a monopole.\\
We can verify the above conclusion from a slighly different point of view. Using
the fact that as pointed out by Dirac the Compton wavelength above is
a region that is minimal in the sense that within it we have the unphysical
Zitterbewegung effects which have to be averaged out, we are lead to a
non commutative geometry\cite{r7}
\begin{equation}
[x,y] \approx 0(l^2)\label{e12}
\end{equation}
and similar relations.
For a non commutative structure we have a strong magnetic field $B$,
which in case of (\ref{e12}) is given by
\begin{equation}
\mu = B l^2 \approx \frac{\hbar c}{e}\label{e13}
\end{equation}
It will be immediately observed that (\ref{e13}) defines the Dirac monopole.\\
Interestingly the monopole given by (\ref{e13}) gives an explanation for the
discreteness of the charge, as is well known which conclusion also follows
from the fact that in equation (\ref{e11}) above the weight $N=1$.

\end{document}